# ELECTROMAGNETIC WAVES WITH NONLINEAR DISPERSION LAW

P. M. Mednis

Novosibirsk State Pedagogical University, Chair of the General and Theoretical Physics, Russia, 630126, Novosibirsk, Viljujsky, 28
e-mail: pmednis@inbox.ru

**Abstract**

Last year physicists in Europe have measured the velocity of the neutrinos particles. They found the neutrinos moving faster than the speed of light in vacuum. This result means that Einstein's relativity principle and its consequences in modern physics need a global additional renovation. In present paper the part of this problem is considered in terms of basic Maxwell's method only. By means of introduction a diffusion like displacement current the new super wave equation was derived, which permits of its solution be described the electromagnetic waves moving some faster than the conventional speed of light in vacuum especially in a gamma ray of a very short wave length region. The unique properties of these waves are that they undergo nonlinear dispersion law, uppermost limit of which is restricted. Discussion of further experimental problems and a number of estimations are given for the macro physic super wave equations also.

## 1. Introduction

The classical microscopic Maxwell's and Lorentz's equations of electrodynamics present the basic components of the Einstein's relativity principle [1]. Let us write them

$$\nabla \cdot \mathbf{B} = 0 \, , \qquad \nabla \times \mathbf{E} = -\frac{1}{c}\frac{\partial \mathbf{B}}{\partial t} \, , \qquad (1)$$

$$\nabla \times \mathbf{B} = \frac{1}{c}\frac{\partial \mathbf{E}}{\partial t} + \frac{4\pi}{c}\mathbf{j} \, , \qquad \nabla \cdot \mathbf{E} = 4\pi\rho \, . \qquad (2)$$

Here we have the *electric field strength vector* $\mathbf{E}$ and the *magnetic induction vector* $\mathbf{B}$. The densities $\mathbf{j}$ and $\rho$ are induced the microscopic electric current density and, respectively, the volume electric charge density which are presented by the expressions

$$\mathbf{j} = \sum_k q_k \mathbf{v}_k \delta(\mathbf{r} - \mathbf{r}_k) \, , \qquad (3)$$

$$\rho = \sum_k q_k \delta(\mathbf{r} - \mathbf{r}_k) \, . \qquad (4)$$

Here the symbol $q_k$ notes the electric charge of *k*-h particle. In addition to the (3) – (4) the vectors $\mathbf{r}$, $\mathbf{r}_k$ are the space and the *k*-h particle radius – vectors, the vector $\mathbf{v}_k = \dot{\mathbf{r}}_k$ is the velocity. The $\delta$ - function is the 3-dimensional Dirac's delta function. The $\rho$ and $\mathbf{j}$ are obeying the charge conservation law

$$\nabla \cdot \mathbf{j} + \frac{\partial \rho}{\partial t} = 0 . \tag{5}$$

The $c$ is the conventional velocity of light constant in vacuum. This is the main goal we are interested in present paper. Indeed, recently European physicists [2] have measured the velocity of the particles called neutrinos. They found the neutrinos moving just faster than the speed of light in vacuum. This unexpected result is enough to call it serious reason that Einstein's relativity principle in all parts of modern physics needs global renovation. So we must begin to revise the basic equations (1) and (2). Our goal is to understand what changes are to be introduced in (1) - (5) and in what sense the Einstein's relativity principle were to be conserved. So, we propose here the new addition to the theory of electrodynamics. They include not the only densities (3) - (4) but additional the like diffusion virtual densities of the charge and current also rising the new components of displacement current. The physical sense this supplements will be seen from the undergoing part of this paper.

## 2. The super wave equation

**2.1. The additional components in the displacement current.** The electric current density (3) and the electric charge density (4) are not the only sources of the electromagnetic field. Long ago, for some reason in nonlinear optics [3] and in course of the field radiation consideration [4] the phenomenological particle was considered may have except the electric charge $e$ the own electric dipole moment $\mathbf{p}(t)$ and the magnetic dipole moment $\boldsymbol{\mu}(t)$ also. Then the selected electric current density $\mathbf{j}_k$ is represented respectively by three components - the transfer, the polarization and the magnetization components. In the physics of semiconductors [5], for example, the different kinds of the diffusion currents are considered. The conventional way of taking into account an additional current carrier and of densities $\mathbf{j}_a$ and $\rho_a$ given is well known. To be sure the conservation law is fulfilled one must write instead of (2) the equations

$$\nabla \times \mathbf{B} = \frac{1}{c}\frac{\partial \mathbf{E}}{\partial t} + \frac{4\pi}{c}(\mathbf{j} + \mathbf{j}_a), \quad \nabla \cdot \mathbf{E} = 4\pi(\rho + \rho_a) \tag{6}$$

with the additional part of the conservation law

$$\nabla \cdot \mathbf{j}_a + \frac{\partial \rho_a}{\partial t} = 0. \tag{7}$$

It will be noted now when a new kind of charge carriers absent but the new components of current $\mathbf{j}_a$ appears at any virtual form at best to correspond with the Maxwell's method we must write an additional electric field $\mathbf{E}_a$ in the displacement current also instead of using the form (6). As a result instead of (6) we have to write

$$\nabla \times \mathbf{B} = \frac{1}{c}\frac{\partial(\mathbf{E}+\mathbf{E}_a)}{\partial t} + \frac{4\pi}{c}(\mathbf{j}+\mathbf{j}_a), \quad \nabla \cdot \mathbf{E} = 4\pi\rho \tag{8}$$

The conservation law one may write now in the alternative form

$$4\pi\nabla \cdot \mathbf{j}_a + \frac{\partial}{\partial t}\nabla \cdot \mathbf{E}_a = 0. \tag{9}$$

First of all these equations permit us find the field $\mathbf{E}_a$ or it's derivative $\frac{\partial \mathbf{E}_a}{\partial t}$ and then define the additional charge density $\rho_a$ by means of using the equation $\nabla \cdot \mathbf{E}_a = 4\pi\rho_a$. So, for the diffusion like electric current of the kind

$$\mathbf{j}_a = -D\mathrm{grad}\rho, \tag{10}$$

where $D$ is the constant coefficient which dimension in SI equal the square of length divided at the time, i.e. $[D] = m^2/s$, we find from (8) and (9) the expression

$$\nabla \cdot (-D\Delta\mathbf{E}) + \frac{\partial}{\partial t}\nabla \cdot \mathbf{E}_a = 0. \tag{11}$$

Here and in further expressions the symbol $\Delta$ is the known Laplacian operator [1]. The main solution of this equation gives us the new component of displacement current and the derivative of the charge $\rho_a$ expressed in terms of electric field $\mathbf{E}$ and density of charge $\rho$ as follows

$$\frac{\partial \mathbf{E}_a}{\partial t} = D\Delta\mathbf{E}, \quad \frac{\partial \rho_a}{\partial t} = D\Delta\rho. \tag{12}$$

For the microscopic dipole diffusion like current of the kind

$$\mathbf{j}'_a = -\kappa \frac{\partial}{\partial t} \operatorname{grad} \rho \tag{13}$$

where $\kappa$ is another constant diffusion coefficient which dimension in SI is equal to the square of length, i.e. $[\kappa] = m^2$, we find from (8) and (9) the expression

$$\frac{\partial}{\partial t} \nabla \cdot (-\kappa \Delta \mathbf{E}) + \frac{\partial}{\partial t} \nabla \cdot \mathbf{E}'_a = 0 \tag{14}$$

As a consequence we have the solution of this equation

$$\mathbf{E}'_a = \kappa \Delta \mathbf{E}, \quad \rho'_a = \kappa \Delta \rho. \tag{15}$$

As a result in summary we may write instead of (8) the equations

$$\nabla \times \mathbf{B} = \frac{1}{c}(1 + \kappa \Delta)\frac{\partial \mathbf{E}}{\partial t} + \frac{1}{c} D \Delta \mathbf{E} + \frac{4\pi}{c}(\mathbf{j} + \mathbf{j}_a + \mathbf{j}'_a), \quad \nabla \cdot \mathbf{E} = 4\pi\rho. \tag{16}$$

Combine (10) and (13) with $\nabla \cdot \mathbf{E} = 4\pi\rho$ we may exclude the $\mathbf{j}_a$ and $\mathbf{j}'_a$ from the left equation of (13) at all giving us the new form of equations

$$\nabla \times \mathbf{B} = (1 - \kappa \nabla \times \nabla \times)\frac{1}{c}\frac{\partial \mathbf{E}}{\partial t} - \frac{1}{c} D \nabla \times \nabla \times \mathbf{E} + \frac{4\pi}{c}\mathbf{j}, \quad \nabla \cdot \mathbf{E} = 4\pi\rho. \tag{17}$$

Also for the current of the particle dipole $\mathbf{p} = e\mathbf{l}$ used in [3,4]

$$\mathbf{j}''_a = \frac{\partial}{\partial t} \mathbf{p} \delta(\mathbf{r} - \mathbf{r}_0(t)) \tag{18}$$

similar consideration yields

$$\mathbf{E}''_a = -(\mathbf{l} \cdot \nabla)\mathbf{E}, \quad \rho''_a = -(\mathbf{l} \cdot \nabla)\rho. \tag{19}$$

On its own account this current call into action the new members in (2)

$$\nabla \times \mathbf{B} = \frac{1}{c}(1 - (\mathbf{l} \cdot \nabla))\frac{\partial \mathbf{E}}{\partial t} + \frac{4\pi}{c}(\mathbf{j} + \mathbf{j}''_a), \quad \nabla \cdot \mathbf{E} = 4\pi\rho. \tag{20}$$

Here also we may exclude $\mathbf{j}''_a$ to get

$$\nabla \times \mathbf{B} = \frac{1}{c}\frac{\partial \mathbf{E}}{\partial t} + \nabla \times \mathbf{l} \times \frac{1}{c}\frac{\partial \mathbf{E}}{\partial t} + \frac{4\pi}{c}\mathbf{j}, \quad \nabla \cdot \mathbf{E} = 4\pi\rho. \tag{21}$$

Because of the latter case have the phenomenological character we return to the microscopic currents (10) and (13) with the main equations (17), (3), (4) and (5). As for Fara-day's equation in (1) to ensure the further wave solutions in absence of any kind of magnetic charge we must generalize this equation for symmetry reason to the form

$$\nabla \times \mathbf{E} = -(1 + \kappa \Delta)\frac{1}{c}\frac{\partial \mathbf{B}}{\partial t} + \frac{1}{c} D \Delta \mathbf{B} \quad , \quad \nabla \cdot \mathbf{B} = 0 \qquad (22)$$

or in the equivalent form

$$\nabla \times \mathbf{E} = -(1 - \kappa \nabla \times \nabla \times)\frac{1}{c}\frac{\partial \mathbf{B}}{\partial t} - \frac{1}{c} D \nabla \times \nabla \times \mathbf{B} \quad , \quad \nabla \cdot \mathbf{B} = 0 \qquad (23)$$

The group of equations (17), (23) and (5) presents the new equations of electrodynamics taking into account the two types of diffusion like currents. It is seen that electromagnetic field in vacuum is characterized not only by the constant $c$ but in addition account must be taken of the pare constant $\kappa$ and $D$. It will be our task below to discuss and estimate them.

**2.2. Field super wave equation.** Exclude from the equations (17) and (23) in turn respectively electric and magnetic fields we may write the

$$\nabla \times \nabla \times \mathbf{E} = -(1 - \kappa \nabla \times \nabla \times)^2 \frac{1}{c^2}\frac{\partial^2 \mathbf{E}}{\partial t^2} + \left(\frac{D}{c}\right)^2 \nabla \times \nabla \times \nabla \times \nabla \times \mathbf{E} - \\ - \frac{4\pi}{c}\left[(1 - \kappa \nabla \times \nabla \times)\frac{\partial \mathbf{j}}{\partial t} + \frac{D}{c}\nabla \times \nabla \times \mathbf{j}\right] \qquad (24)$$

$$\nabla \times \nabla \times \mathbf{B} = -(1 - \kappa \nabla \times \nabla \times)^2 \frac{1}{c^2}\frac{\partial^2 \mathbf{B}}{\partial t^2} + \left(\frac{D}{c}\right)^2 \nabla \times \nabla \times \nabla \times \nabla \times \mathbf{B} + \frac{4\pi}{c}\nabla \times \mathbf{j} \qquad (25)$$

For the free of charge space we get the equations

$$\Delta\left(1 + \left(\frac{D}{c}\right)^2 \Delta\right)\mathbf{E} - (1 + \kappa \Delta)^2 \frac{1}{c^2}\frac{\partial^2 \mathbf{E}}{\partial t^2} = 0 \quad , \quad \nabla \cdot \mathbf{E} = 0 \qquad (26)$$

$$\Delta\left(1 + \left(\frac{D}{c}\right)^2 \Delta\right)\mathbf{B} - (1 + \kappa \Delta)^2 \frac{1}{c^2}\frac{\partial^2 \mathbf{B}}{\partial t^2} = 0 \quad , \quad \nabla \cdot \mathbf{B} = 0 \qquad (27)$$

For the further development such equations are named as *super wave equations* (SWE). In the free space SWE permits the plane wave solutions

$$\mathbf{E} = \mathbf{E}_0 e^{i(\mathbf{kr} - \omega t)} \quad , \quad \mathbf{B} = \mathbf{B}_0 e^{i(\mathbf{kr} - \omega t)} \qquad (28)$$

with the constant complex amplitudes $\mathbf{E}_0$, $\mathbf{B}_0$, which are perpendicular to the wave vector $\mathbf{k}$, i. e. the transverse wave conditions $\mathbf{k} \cdot \mathbf{E}_0 = 0$ and $\mathbf{k} \cdot \mathbf{B}_0 = 0$ must be satisfied. The symbol ω notes a circle frequency in course of time process. More over this frequency depends on the additional constants κ and $D$. The dispersion law is

$$\omega = \pm kc \frac{\sqrt{1 - \left(\frac{D}{c}\right)^2 k^2}}{\left|1 - \kappa k^2\right|}. \tag{29}$$

Consider now the score of the possible solutions and their graphical presentation with the help of MathCad computer program. In values region of wave vectors $k \leq c/D$ the frequency ω is real, so the wave solution (28) with arbitrary sign of (29) exists. If the condition $k > c/D$ takes place the frequency ω became imaginary. A physical damped solution (28) in this case must be constructed closely with choose of the sign of (29). As we can see from denominator of (29) the singular point exists if the wave vector obeys the condition $k = 1/\sqrt{\kappa}$. For the not damped regular wave solution (28) the condition $k \leq c/D < 1/\sqrt{\kappa}$ realized. In this case the dependency of $\omega D/c^2$ on $D\kappa/c$ for the $\kappa c^2/D^2 = 0.5$ as example is seen at the Fig. 1.

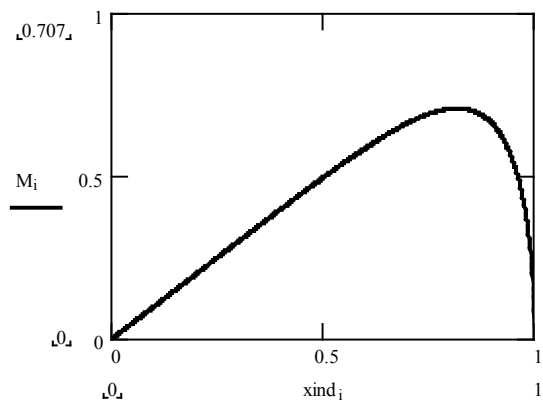

Figure 1: The counter plot of the $\omega D/c^2$ versus the $Dk/c$ for the $\kappa c^2/D^2 = 0.5$

The phase speed $v_{ph}$ of this wave is given as usual by expression

$$V_{ph} = \frac{\omega}{k} = c \frac{\sqrt{1-\left(\frac{D}{c}\right)^2 k^2}}{\left|1-\kappa k^2\right|} \quad . \tag{30}$$

The dependency of $v_{ph}/c$ on $Dk/c$ for the same $\kappa c^2/D^2 = 0.5$ is seen at the Fig. 2.

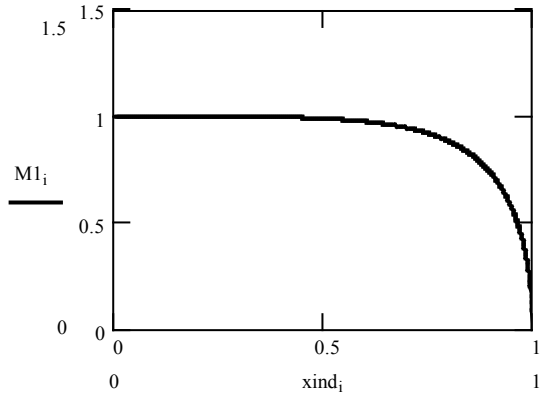

Figure 2: The counter plot of the $v_{ph}/c$ versus the $Dk/c$ for the $\kappa c^2/D^2 = 0.5$

For the not damped wave solution (28) under condition $k < 1/\sqrt{\kappa} \leq c/D$ the point of singularity presents inside the region. In this case the dependency of $\omega D/c^2$ on $Dk/c$ for the $\kappa c^2/D^2 = 1.5$ as example is seen at the Fig. 3.

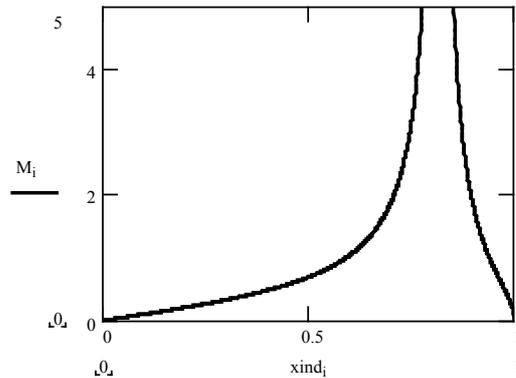

Figure 3: The counter plot of the $\omega D/c^2$ versus the $Dk/c$ for the $\kappa c^2/D^2 = 1.5$

The dependency of $v_{ph}/c$ on $Dk/c$ for the same $\kappa c^2/D^2 = 1.5$ is seen at the Fig. 4.

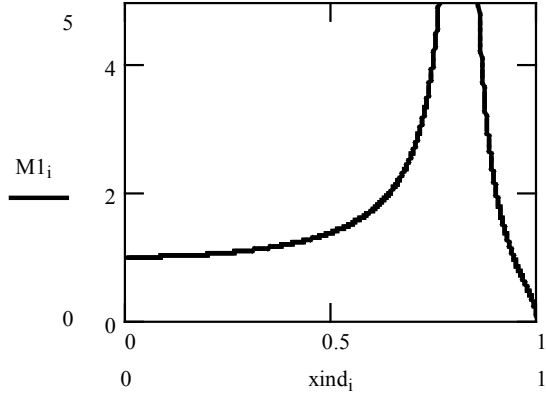

Figure 4: The counter plot of the $v_{ph}/c$ versus the $Dk/c$ for the $\kappa c^2/D^2 = 1.5$

It seems the inherent damped waves and the interruption in the spectrum for the classical vacuum space are to be the nonphysical solutions. To avoid such solutions we must demand the condition $D/c = \sqrt{\kappa}$ to be taken place. For this case the dispersion law is in form

$$\omega = \frac{kc}{\sqrt{1-k^2/k_0^2}}, \quad k_0 = \frac{1}{\sqrt{\kappa}} = \frac{2\pi}{\lambda_0}. \tag{31}$$

The dependency of $\omega/k_0 c$ on $k/k_0$ is seen at the Fig. 5.

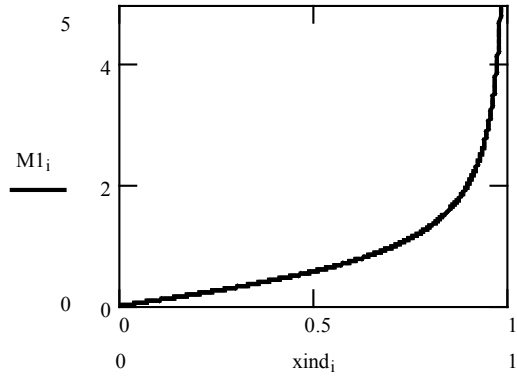

Figure 5: The counter plot of the $\omega/k_0 c$ versus the $k/k_0$

The phase speed $v_{ph}$ in this case is

$$v_{ph} = \frac{\omega}{k} = \frac{c}{\sqrt{1-k^2/k_0^2}}. \tag{32}$$

The dependency of $v_{ph}/c$ on $k/k_0$ is seen at the Fig. 6.

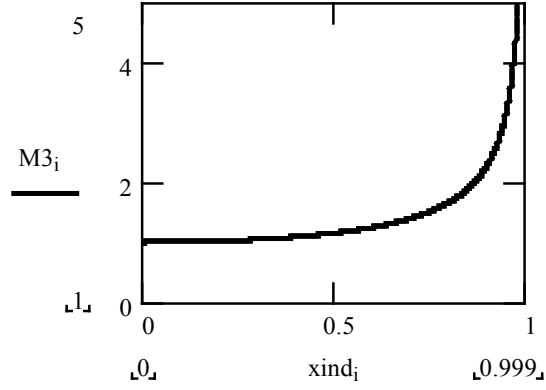

Figure 6: The counter plot of the $v_{ph}/c$ versus the $k/k_0$

In addition as another option to avoid any damped waves in a broad manner we must suggest also that the constant $D = 0$ in (10) at all. Then instead of (31) and (32) we must write respectively

$$\omega = \frac{kc}{\left|1 - k^2/k_0^2\right|}, \quad k_0 = \frac{1}{\sqrt{\kappa}} = \frac{2\pi}{\lambda_0} \tag{33}$$

and

$$v_{ph} = \frac{\omega}{k} = \frac{c}{\left|1 - k^2/k_0^2\right|}. \tag{34}$$

solely under condition $k < k_0$. The graphics of that function are similar in a sense to of Fig. 5 and Fig. 6 respectively. What case is realized really must be solved by a future experiment. My own preference to be related to the last case because of some its natural simplicity. Indeed, the maximal value of wave vector $k_0$ exists because of the minimal spatial parameter $\lambda_0$ or a called minimal space length exists and broadly are discussed until now. The further part of this paper is considering under condition that $D = 0$.

**2.3. Field potentials super wave equations.** From the (22) by means of standard method we derive the connections of fields and potentials

$$\mathbf{B} = \nabla \times \mathbf{A}, \quad \mathbf{E} = -(1 + \kappa\Delta)\frac{1}{c}\frac{\partial \mathbf{A}}{\partial t} - \nabla\varphi. \tag{35}$$

Here respectively **A** and φ are the vector and scalar potentials. Then from (16) follow the equations for potentials

$$\Delta \mathbf{A} - (1+\kappa\Delta)^2 \frac{1}{c^2}\frac{\partial^2 \mathbf{A}}{\partial t^2} = \nabla\left[\nabla\cdot\mathbf{A} + \frac{1}{c}(1+\kappa\Delta)\frac{\partial\varphi}{\partial t}\right] - \frac{4\pi}{c}\left(\mathbf{j} - \kappa\nabla\frac{\partial\rho}{\partial t}\right), \qquad (36)$$

$$\Delta\varphi - (1+\kappa\Delta)^2 \frac{1}{c^2}\frac{\partial^2 \varphi}{\partial t^2} = -\frac{1}{c}(1+\kappa\Delta)\frac{\partial}{\partial t}\left[\nabla\cdot\mathbf{A} + \frac{1}{c}(1+\kappa\Delta)\frac{\partial\varphi}{\partial t}\right] - 4\pi\rho. \qquad (37)$$

As well known the potentials **A** and φ are not unique values characterizing the fields. The *gradient* or *calibration* transformation in (35) to another pair of potentials $\widetilde{\mathbf{A}}$ and $\widetilde{\varphi}$ with any arbitrary function ψ should be written in the form

$$\mathbf{A} = \widetilde{\mathbf{A}} - \nabla\psi, \quad \varphi = \widetilde{\varphi} + \frac{1}{c}(1+\kappa\Delta)\frac{\partial\psi}{\partial t}. \qquad (38)$$

If needed this property permits us to simplify (36) and (37). So in particular new Lorentz's calibration is presented now in the form

$$\nabla\cdot\mathbf{A} + \frac{1}{c}(1+\kappa\Delta)\frac{\partial\varphi}{\partial t} = 0. \qquad (39)$$

However the Coulomb's calibration became unchanged

$$\nabla\cdot\mathbf{A} = 0. \qquad (40)$$

## 3. Further more

**3.1. How we can verify the theory**. As a result of our theory for the case $D=0$ also we are thus led to the concept that the speed of light in vacuum is no more the conventional constant value $c$. The dependence of the speed of light versus the relative wave number $k/k_0$ is given by (34). The real value of $k_0$ is unknown now. In the full lot of early experiments it has been showed that the speed of light in vacuum is the constant value $c$ because I think of it was the constant by the definition mostly. Indeed, this result may mean no more that the value of $k_0$ is very large or as seen from (33) the value of $\lambda_0$ is a very small value also. To estimate the $\lambda_0$ let us take the relative speed difference of the same order as in [2], for example, i. e.

$$\frac{V_{ph} - c}{c} = 2.5 \cdot 10^{-5}.  \quad (41)$$

Then for the very high energetic gamma rays which length of wave is equal $\lambda = 10^{-14} m$ we get for the value $\lambda_0 = 5 \cdot 10^{-17} m$. This is the value of the order of the *fundamental length*, which estimated value supposedly lies now in this region and less indeed. From (33) it follows that $k_0 = 1.25 \cdot 10^{17} m^{-1}$. Hence the dependence (34) can be identified for the very short wave length gamma rays region only. For macrophysics case of fields we must consider the new equations (16) – (22) or (17) – (23) as the averaged one by microscopic initial states with any distribution function [4]. Up to visual optic region we may hope to check the additional current effects by means of polarizations effects only because of a small new addition in boundary conditions for electric field strength vector **E** depending on $\kappa$.

**3.2. The ways and means to renovate the Einstein's relativistic principle**. The coordinates and time transformation of equations (36) and (37) are the more so interesting for us now because of the transformation from one inertial system to another inertial system of references. In accordance with the principle of relativity the Maxwell's equation (1) and (2) are unchanged in form under the linear Lorenz's transformation [1]. Let be the group of spatial coordinates $x, y, z$ and the time $t$ are given in an inertial system of references $K$ in which the equations (1) and (2) are given. Let be another group of spatial the coordinates $x', y', z'$ and the time $t'$ are given in an inertial system of references $K'$ which is moving with the constant speed $V$ relatively to a system of references $K$ in the positive direction of $x$ axis so the all directions of axis $x, y, z$ and of axis $x', y', z'$ remains parallel to each other. For this case the Lorenz's transformations are

$$x = \frac{x' + Vt'}{\sqrt{1 - \frac{V^2}{c^2}}}, \ y = y', \ z = z', \ t = \frac{t' + \frac{V}{c^2}x'}{\sqrt{1 - \frac{V^2}{c^2}}}. \quad (42)$$

The reverse transformations are found from (42) if the all changes of quantities $x, y, z, t \leftrightarrow x', y', z', t'$ plus the change $V \leftrightarrow -V$ are realized respectively. For beginning in our case all we can write now for equations (36) and (37) the symbolic transformations in the form

$$x = x(x', t'), \quad y = y', \quad z = z', \quad t = t(x', t') \quad , \tag{43}$$

$$x' = x'(x, t), \quad y' = y, \quad z' = z, \quad t' = t'(x, t) \quad . \tag{44}$$

To transform the potentials in the equations (36) and (37) we must combine the vector and scalar potentials and change variables as follows

$$A_x = \gamma_{11} A'_x + \gamma_{12} \varphi', \quad A_y = A'_y, \quad A_z = A'_z, \tag{45}$$

$$\varphi = \gamma_{21} A'_x + \gamma_{22} \varphi'. \tag{46}$$

In this equations the parameters $\gamma_{ij}$ ($i, j = 1, 2$) are not depending on the coordinates and the time. The primed potential $\mathbf{A}'$ and $\varphi'$ in these equations are depending on $x', y', z', t'$. The starting potentials on the left side of (36) and (37) are depending on $x, y, z, t$. Obviously the reverse transformation of (45) and (46) may be found easily. Substitute the (45) and (46) in the expression standing in square bracket of (36) and (37) equations. Then we can find that this expression is invariant, i.e.

$$\nabla \cdot \mathbf{A} + (1 + \kappa \Delta) \frac{1}{c} \frac{\partial \varphi}{\partial t} = \nabla' \cdot \mathbf{A}' + (1 + \kappa \Delta') \frac{1}{c} \frac{\partial \varphi'}{\partial t'} = \text{inv} \quad , \tag{47}$$

if the rules of transformation of the differentials are defined as follows

$$\frac{\partial}{\partial x'} = \gamma_{11} \frac{\partial}{\partial x} + \gamma_{21} (1 + \kappa \Delta) \frac{1}{c} \frac{\partial}{\partial t}, \quad \frac{\partial}{\partial y'} = \frac{\partial}{\partial y}, \quad \frac{\partial}{\partial z'} = \frac{\partial}{\partial z}, \tag{48}$$

$$(1 + \kappa \Delta') \frac{1}{c} \frac{\partial}{\partial t'} = \gamma_{12} \frac{\partial}{\partial x} + \gamma_{22} (1 + \kappa \Delta) \frac{1}{c} \frac{\partial}{\partial t}. \tag{49}$$

The reverse transformations of these differentials are written respectively in form

$$\frac{\partial}{\partial x} = \frac{1}{\gamma_{11}\gamma_{22} - \gamma_{12}\gamma_{21}} \left[ \gamma_{22} \frac{\partial}{\partial x'} - \gamma_{21}(1 + \kappa\Delta') \frac{1}{c} \frac{\partial}{\partial t'} \right], \quad \frac{\partial}{\partial y} = \frac{\partial}{\partial y'}, \quad \frac{\partial}{\partial z} = \frac{\partial}{\partial z'}, \tag{50}$$

$$(1+\kappa\Delta)\frac{1}{c}\frac{\partial}{\partial t} = -\frac{1}{\gamma_{11}\gamma_{22} - \gamma_{12}\gamma_{21}}\left[\gamma_{12}\frac{\partial}{\partial x'} - \gamma_{11}(1+\kappa\Delta')\frac{1}{c}\frac{\partial}{\partial t'}\right]. \quad (51)$$

In addition, special account must be taken of Laplacians on the left sides of (49) and (51), they were expressed by means of (48) and (50) respectively. Also we suppose that both reverse operators $(1+\kappa\Delta')^{-1}$ and $(1+\kappa\Delta)^{-1}$ exist. After substitution (45) and (46) in (36) and (37) we can see they are transformed to invariant form if the following conditions for $\gamma_{ij}$ takes place

$$\gamma_{11} = \gamma_{22}, \; \gamma_{12} = \gamma_{21}, \; \gamma_{11}^2 - \gamma_{12}^2 = 1. \quad (52)$$

However the explicit sight of the transformation functions $x = x(x',t')$ and $t = t(x',t')$ may not be found so easily as we want. I stop further development of searching the solutions for a variety of reasons. First of all the functions $x = x(x',t')$ and $t = t(x',t')$ are expected to be the nonlinear functions it may be of a linear combinations of $x'$ and $t'$ resulting some technical difficulties for defining the constants $\gamma_{ij}$ and the integration constants also. Besides that to renovate Einstein's relativistic principle by means of the simple exchange of Lorenz's transformations (42) by any new transformations only does not possible in principle. The main reason is that equations (36) and (37) are to be written in a new invariant form to define the called *action* and then to find a new equation of motions for charged particle using the *principle of minimal action* [6]. Unfortunately this procedure needs intra generalization to be applied to the case of derivatives of higher than second order in the equations (36) and (37). It will be very interesting to solve both these problems stated due to the full scope of interests mainly for the wide range of new macroscopic applications. Naturally the starting point of a solution may differ from the suggested one.

## 4. References

**1.** J. D. Jackson. Classical Elecrodynamics, 3rd ed. Wiley, New York, 1999.

**2.** Measurement of the neutrino velocity with the OPERA detector in the CNGS beam. Arxiv.org/abs/1109.4897

**3.** V.N. Genkin, P.M. Mednis. About the own electric dipole and quasi-orbital magnetic moments of a conductivity electron. Izv. Vuzov, GGU, Radio-fizika, **10** (1967), 585.

**4.** V.L. Ginzburg. The theoretical physics and astrophysics. Moscow, Nauka, 1981.

**5.** C.F. Klingshirn. Semiconductor optics. 3$^{rd}$ ed. Springer, Berlin 2007.

**6.** L.D. Landau, E.M. Lifshitz. The classical theory of fields. Oxford, Butterworth – Heinemann, 1995.